\documentclass[aps,prb,amsmath,amssymb,twocolumn,superscriptaddress,preprintnumbers,showpacs,longbibliography]{revtex4-2}
\pdfoutput=1
\bibpunct{[}{]}{,}{n}{}{}

\usepackage[utf8]{inputenc}
\usepackage{bm,latexsym,mathrsfs,enumerate}
\usepackage[dvipsnames]{xcolor}
\usepackage{mathtools}
\usepackage{makecell}
\usepackage{multirow}
\usepackage[breaklinks=true,unicode=true,urlcolor = blue,colorlinks = true,citecolor = blue,linkcolor = blue]{hyperref}
\usepackage{graphicx}
\usepackage{ragged2e}
\usepackage[export]{adjustbox}

\renewcommand{\vec}[1]{\bm{#1}}
\usepackage[final]{pdfpages}
\usepackage{tikz}
\makeatletter
\AtBeginDocument{\let\LS@rot\@undefined}
\makeatother

\begin{document}

\title{Theory of spin-polarized high-resolution electron energy-loss spectroscopy from nonmagnetic surfaces with a large spin--orbit coupling}

\author{Khalil Zakeri}
\email{khalil.zakeri@kit.edu}

\affiliation{Heisenberg Spin-dynamics Group, Physikalisches Institut, Karlsruhe Institute of Technology, Wolfgang-Gaede-Str.\ 1, D-76131 Karlsruhe, Germany}

\author{Christophe Berthod} 

\affiliation {Department of Quantum Matter Physics, University of Geneva, 1211 Geneva, Switzerland}

\begin{abstract}

The scattering theory of low-energy (slow) electrons has been developed by Evans and Mills [\href{https://doi.org/10.1103/PhysRevB.5.4126}{Phys. Rev. B \textbf{5}, 4126 (1972)}]. The formalism is merely based on the electrostatic Coulomb interaction of the scattering electrons with the charge-density fluctuations above the surface and can describe most of the interesting features observed in the high-resolution electron energy-loss spectroscopy experiments. Here we extend this theory by including the spin--orbit coupling in the scattering process. We discuss the impact of this interaction on the scattering cross-section. In particular, we discuss cases in which a spin-polarized electron beam is scattered from nonmagnetic surfaces with a strong spin--orbit coupling. We show that under some assumptions one can derive an expression for the scattering cross-section, which can be used for numerical calculations of the spin-polarized spectra recorded by spin-polarized high-resolution electron energy-loss spectroscopy experiments.

\end{abstract}
\maketitle


\section{Introduction}

Experimental techniques based on particle scattering belong to the most powerful tools for probing and investigation of elementary collective excitations in solids. The type and the excitation probability of such collective modes is determined by the type and the strength of the microscopic physical interaction describing the coupling between the scattering particle and the sample.
In the case of neutrons or neutral atoms the interaction is mainly between the incoming particle and the atomic nuclei. If the particle possesses a spin (e.g., in the case of
neutrons) the spin--spin interaction shall also be taken into account \cite{Chatterji2006, Squires2012, Berthod2018}. Since electrons are charged particles, when a beam of low-energy (slow) electrons is scattered from a surface, it interacts with the charge-density fluctuations near the surface region and can couple to collective charge excitations. The interaction is of Coulomb nature and hence is strong and long range. This is the central mechanism behind probing collective excitations associated with the charge degree of freedom, e.g., phonons and plasmons (or any hybrid mode of these two) using low-energy (slow) electrons in high-resolution electron energy-loss spectroscopy (HREELS) experiments \cite{Ibach1982}. The spectral function probed by these experiments is, in fact, directly proportional to the dynamic charge response of the sample.

In the electron scattering experiments, e.g., HREELS one usually distinguishes two regimes, i.e., dipole and impact scattering. Dipole scattering refers to scattering geometries in the vicinity of the specular reflection. This is commonly referred to as the dipolar lobe and exhibits a very narrow angular distribution. In this regime, the cross-section can be calculated without knowing the microscopic details of the sample \cite{Mills1975}. However, at large deflection angles a detailed and microscopic knowledge of the sample is required in order to describe the scattering process. The scattering at large deflection angles outside the dipolar lobe is commonly referred to as impact scattering. The angular distribution  of impact scattering is very broad and the scattering intensity is by several orders of magnitude smaller than that of the dipole scattering.

The theory of low-energy electron scattering has been developed by Evans and Mills long time ago \cite{Evans1972, Mills1975, Ibach1982}. In that formalism the spin  degree of freedom of electrons and the relativistic effects have not been taken into account. Perhaps this is due to the fact that there has been no demand and inspiration from the experimental side. The conventional  HREELS experiments have  been performed using an unpolarized electron beam. Hence, the spin-dependent effects could not be measured in those experiments. The focus has mainly been on the investigation of surface phonons \cite{Ibach1982}. 
The scattering cross-section obtained within that framework has successfully been used for the numerical calculations of the HREELS spectra using the numerical scheme developed by Lucas and co-workers \cite{Sunjic1971, Lucas1972, Lambin1990}.
The theory has recently been extended to the case in which electrons transfer a momentum equal to a reciprocal-lattice vector to the sample during the scattering process \cite{Vig2017}. It has been shown that the formalism of Evans and Mills is also valid for such cases \cite{Vig2017}. 
In addition to their charge, electrons possess also a spin and hence spin-polarized electron scattering experiments can provide valuable information on spin-dependent excitations. In the traditional spin-polarized electron energy-loss spectroscopy experiments the energy and momentum resolution have not been sufficient to resolve the low-energy collective excitations and their spin dependence. The central idea of those experiments has mainly been to study concepts like Stoner excitations at magnetic surfaces or inter-/intra-band transitions, both of which take place at rather high energy losses (typical a few eV) \cite{Kirschner1984, Hopster1984, Kirschner1985, Kirschner1985a, Hopster1985}.

Here we revisit the theory of low-energy electron scattering and extend it to a case in which a spin-polarized electron beam is scattered from a surface with a large spin--orbit coupling (SOC). We restrict ourselves to the dipole scattering regime and nonmagnetic surfaces.  We will show that the main advantage of our approach is that it does not require the microscopic details of the sample.

The spin-polarized scattering experiments become of importance in two classes of materials, i.e., (i) magnetically-ordered solids and (ii) solids with a large SOC. In the former class, in particular in the case of ferromagnetic metals, the exchange process is of great importance and it is the underlying mechanism of spin-flip excitations, e.g., magnons as quanta of spin waves and single-particle Stoner excitations \cite{Kirschner1984, Hopster1984, Kirschner1985, Kirschner1985a, Hopster1985, Vollmer2003, Prokop2009, Zakeri2010, Zhang2011a, Zakeri2013a, Zakeri2012, Zakeri2013, Zakeri2014, Zakeri2018a}. The mechanism leading to these excitations is the so-called exchange scattering, which is of Coulomb nature.
The fundamental physical mechanisms behind inelastic electron scattering leading to spin-flip excitations are inherently different from those of inelastic neutron scattering, even though both lead to very similar results. Since electrons are indistinguishable particles, in electron scattering processes the exchange mechanism plays an important role. On the contrary, in neutron scattering experiments the type of the interaction that is important is the magnetic spin--spin interaction between the neutron's magnetic moment and the magnetic moment of the unit cell, which leads to the magnon excitations. In the electron scattering experiments one may imagine the magnon excitation process within the following classical picture. Let us assume that an electron with the spin parallel to that of the minority electrons of the ferromagnetic sample is incident onto the surface.
Suppose that the incident electron fills an unoccupied state above the Fermi-level while a majority electron from an occupied state below the Fermi-level is scattered out. This exchange process leads to a virtual spin flip and consequently an electron--hole pair with total angular momentum 1 $\hbar$ across the Fermi-level ($\hbar$ is the reduced Planck's constant). The excitation of magnons by spin-polarized electrons is based on such a process. The scattering itself is elastic and the observed energy loss (or gain) of electrons is due to the fact that the ejected electron originates from a lower (or higher) energy level of the excited solid. Such a process is mediated by exchange interaction that is of Coulomb nature and no explicit spin--spin interaction is needed to be taken into account. The magnon excitation process is very fast and occurs within a few attoseconds. For an extended discussion the reader is referred to Refs.~\cite{Zakeri2013, Zakeri2014}. Based on the arguments discussed above and since we do not deal with magnetic surfaces here, we do not consider any spin--spin interaction in our formalism \footnote{The electrons also experience a purely magnetic scattering, much weaker than Coulomb scattering, due to their magnetic moment. The typical energy scale for this magnetic scattering in a volume $a_0^3$ is $(g_e\mu_{\mathrm{B}})^2\mu_0/a_0^3=0.02$~eV, with $a_0$ the Bohr radius, $g_e\approx-2$, $\mu_{\mathrm{B}}$ the Bohr magneton, and $\mu_0$ the vacuum permeability. This must be compared with the energy scale $e^2/(\epsilon_0a_0^3q^2)$ for Coulomb scattering [see Eq.~(\ref{Eq:HartreeMatrixElement})], which is 342~eV for $q=1/a_0$.}.

It has been realized that when a beam of spin-polarized slow electrons is elastically scattered from a surface with a large SOC, the scattering intensity can exhibit a spin asymmetry, depending on the relative orientation of the spin of the incoming beam with respect to the scattering plane \cite{Kessler1985, Kirschner1985, Feder1986, Wang1979}. The effect is a direct consequence of SOC \cite{Wang1979, McRae1979, McRae1981, Kessler1985, Kirschner1985, Feder1986, Yu2007, Burgbacher2013, Vasilyev2015}. The asymmetry caused by SOC is also present in the inelastic part of the scattering. However, there are still several unresolved fundamental questions. (i) How is the asymmetry of the inelastic part of the scattering connected to that of the elastic scattering in high-resolution experiments? (ii) How does SOC influence the scattering cross-section? (iii) Does the observed spin asymmetry depend on the energy loss and momentum transfer?

We discuss the influence of the SOC term on the scattering cross-section and the SOC-induced spin asymmetry for both the elastically and inelastically scattered electrons. The results are particularly important to describe spin-polarized high-resolution electron energy-loss spectroscopy (SPHREELS) experiments performed close to the specular reflection on nonmagnetic surfaces with a strong SOC. We note that a fully relativistic theory shall take into account both SOC and the exchange interaction and hence should describe all the details of the spin-polarized spectra, regardless of the magnetic state of the sample. However, developing such a theory requires (i) a sophisticated description of the exchange process within both the dipole and impact scattering regimes and (ii) all the microscopic details of the sample e.g., details of the geometrical structure near the surface, details of the electronic structure as well as the electronic states involved in the scattering process. 
Although state of the art first-principles calculations may provide the latter requirement, it is not easy to include the exchange process in the scattering event, in particular in the inelastic part and in the impact regime.

\begin{figure}[t!]
	\centering
	\includegraphics[width=0.95\columnwidth]{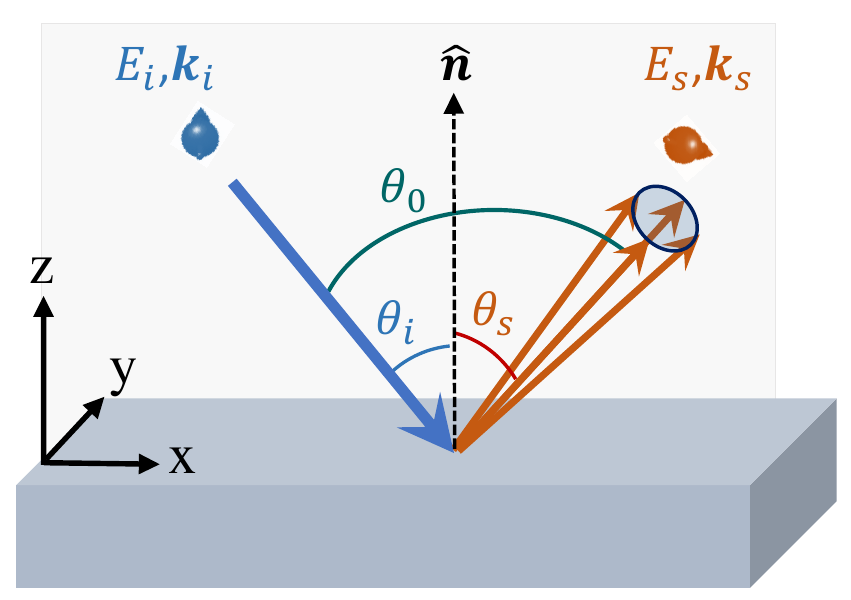}
	\caption{A schematic representation of the scattering geometry used to describe the scattering process. A spin-polarized beam with a given polarization vector is incident onto the sample surface. The incident energy and wavevector are denoted by $E_i$ and $\vec{k}_i$, respectively. The energy and the wavevector after the scattering event are indicated by $E_s$ and $\vec{k}_s$, respectively. The laboratory frame is depicted in the left corner with coordinates $x$, $y$, and $z$. The sample surface is placed in the $x$--$y$ plane at $z=0$. The incident and outgoing angles are called $\theta_i$ and $\theta_s$, respectively. The total scattering angle is $\theta_0$. The unit vector of surface normal is indicated by $\hat{\vec{n}}$. The scattering plane is shown by the shaded area.}
	\label{Fig:Geometry}
\end{figure}

\section{Basic theory}

We consider a case in which a well-defined spin-polarized electron beam with a given energy $E_i$ and momentum $\vec{k}_i$ is scattered from a surface. The energy and momentum of the electron after the scattering process are $E_s$ and $\vec{k}_s$, respectively.
The geometry of such a scattering event is schematically drawn in Fig.~\ref{Fig:Geometry}. The angles $\theta_i$ and $\theta_s$ denote the incident and scattered angles, respectively, and $\theta_0=\theta_i+\theta_s$ indicates the total scattering angle. Assuming that the polarization vector of the incident beam is an arbitrary but known vector in the laboratory frame, one can define its components in the Cartesian coordinates shown in Fig.~\ref{Fig:Geometry}, using its polar and azimuthal angles.

We first define the many-body states of the sample with the energies $E_m$ and $E_n$ as $|m\rangle$ and $|n\rangle$, respectively. The total charge-density operator is given by $\hat{\rho}(\vec{R})$ and includes both negative, e.g., electrons as well as positive charges, e.g., protons. The three dimensional position vector $\vec{R}$ can be decomposed into an in-plane $\vec{r}$ and an out-of-plane component $z$ via $\vec{R}=(\vec{r},z)$. Here $z$ represents the coordinate normal to the surface and the sample is placed in the $x$--$y$ plane at $z<0$ (the surface is located at $z=0$, see Fig.~\ref{Fig:Geometry}). The charge operator $\hat{\rho}(\vec{R})$ acts on the many-body states of the sample with matrix elements $\langle n|\hat{\rho}(\vec{R})|m\rangle$.

The general definition of the differential scattering cross-section is given by \cite{Berthod2018}
\begin{multline}\label{Eq:crossgeneral}
	\frac{d^2S}{d\Omega d\hbar\omega}=\left(\frac{2\pi}{\hbar}\right)^4m_e^2\frac{k_s}{k_i}
	\frac{1}{Z}\sum_{mn}e^{-E_m/k_{\mathrm{B}}T}\times\\
	|\langle n,s|\hat{T}(E_m+E_i)|m,i\rangle|^2\delta(E_m+E_i-E_n-E_s),
\end{multline}
where $d\Omega$ represent the solid angle of the scattering, $m_e$ is the mass of electron, $k_i=|\vec{k}_i|$ and $k_s=|\vec{k}_s|$ represent the norms of the wavevectors of the incident and scattered electron, respectively, $Z=\sum_m e^{-E_m/k_{\mathrm{B}}T}$ is the partition function, $k_{\mathrm{B}}$ is the Boltzmann constant, and $\hbar\omega=E_i-E_s$ represents the energy loss of electrons during the scattering process. We are mostly interested in cases in which $E_s \approx E_i$ and $\hbar\omega \ll E_i$. $\hat{T}(E)$ is the many-body t-matrix given by $\hat{T}(E)=\hat{V}+\hat{V}(E+i0-\hat{H})^{-1}\hat{T}(E)$ with $\hat{H}$ the Hamiltonian of the sample and $\hat{V}$ the interaction energy of the incident electron with the sample. Using the first-order (Born) approximation, the many-body t-matrix reduces to the interaction energy $\hat{T}(E)\approx\hat{V}$. Now if one ignores the exchange-correlation potential, the interaction energy of the incident electron with the sample can be written as the sum of the Coulomb (Hartree) and SOC terms $\hat{V}=\hat{V}_{\mathrm{H}}+\hat{V}_{\mathrm{SOC}}$. $\hat{V}_{\mathrm{H}}$ is the Hartree energy, i.e., the electrostatic interaction with the charge density $\hat{\rho}(\vec{R})$
\begin{equation}\label{Eq:HartreeEnergy}
	\hat{V}_{\mathrm{H}}(\vec{R})=\sigma_0\,e\int d\vec{R}'\,\frac{\hat{\rho}(\vec{R}')}
	{4\pi\epsilon_0|\vec{R}-\vec{R}'|}.
\end{equation}
$\sigma_0=\footnotesize\begin{pmatrix}1&0\\0&1\end{pmatrix}$ is the identity matrix in spin space. The Hartree potential is diagonal in spin space and
conserves the spin of the electron during the scattering process. $\hat{V}_{\mathrm{SOC}}$ is the spin--orbit interaction given by
\begin{equation}\label{Eq:SOCEnergy}
	\hat{V}_{\mathrm{SOC}}(\vec{R})=\frac{e\hbar}{4m_e^2c^2}\,\vec{\sigma}\cdot
	\big[\hat{\vec{E}}(\vec{R})\times\vec{p}\big].
\end{equation}
$\vec{\sigma}$ is the vector of Pauli matrices with the components $\sigma_x=\footnotesize\begin{pmatrix}0&1\\1&0\end{pmatrix}$, $\sigma_y=\footnotesize\begin{pmatrix}0&-i\\i&0\end{pmatrix}$, and $\sigma_z=\footnotesize\begin{pmatrix}1&0\\0&-1\end{pmatrix}$. $\hat{\vec{E}}(\vec{R})=-\vec{\nabla}\hat{V}_{\mathrm{H}}(\vec{R})/e$ is the electric field due to the charge distribution $\hat{\rho}(\vec{R})$, and $\vec{p}=-i\hbar\vec{\nabla}$ is the momentum operator.

According to Eq.~(\refeq{Eq:crossgeneral}), one needs the square of the scattering matrix elements $|\langle n,s|\hat{V}|m,i\rangle|^2$ in order to calculate the scattering cross-section in the Born approximation. To do so, one first needs to define the wavefunctions of the incident and scattered beams.
These wavefunctions are the solutions of the single-particle Schr\"{o}dinger equation with scattering boundary conditions in the half space above the surface $z>0$. Following Refs.~\cite{Evans1972, Mills1975, Vig2017}, we assume a flat surface acting as a hard wall and preventing the incident electrons from penetrating the half space $z<0$. We furthermore generalize these previous works by allowing the surface to display spin-dependent reflection, as may arise due to SOC in the sample or any other mechanism. Under these assumptions, the wavefunctions of the incident $\psi_i(\vec{R})$ and scattered $\psi_s(\vec{R})$ electrons can be written as
\begin{subequations}\label{Eq:ScatteringStates}\begin{align}
	\psi_i(\vec{R})&=N_ie^{i\vec{k}_i\cdot\vec{r}}\left(\vec{i}e^{ik_i^zz}
	+\mathcal{R}\vec{i}e^{-ik_i^zz}\right)\Theta(z)\\
	\psi_s(\vec{R})&=N_se^{i\vec{k}_s\cdot\vec{r}}\left(\vec{s}e^{ik_s^zz}
	+\mathcal{R}\vec{s}e^{-ik_s^zz}\right)\Theta(z).
\end{align}\end{subequations}
Here $N_i$ and $N_s$ are normalization factors. $\Theta(z)$ is the Heaviside step function and ensures that the electrons do not cross the surface. $\vec{i}$ and $\vec{s}$ are two-component vectors representing the initial and final states in some spin-1/2 basis and $\mathcal{R}$ is the reflection matrix expressed in the same basis. It is a $2\times2$ matrix with elements representing the spin-dependent reflection coefficients.

Equations~(\ref{Eq:crossgeneral})--(\ref{Eq:ScatteringStates}) define the spin-resolved scattering cross-section. In the next section, we provide explicit expressions for the Hartree and SOC matrix elements.

\section{Scattering matrix elements}

It is important to mention that while evaluating Eq.~(\ref{Eq:crossgeneral}) both Hartree $\hat{V}_{\mathrm{H}}$ and spin--orbit $\hat{V}_{\mathrm{SOC}}$ terms should be considered in the matrix element $|\langle n,s|\hat{V}|m,i\rangle|^2$. This is a tedious task and leads to long expressions due to the cross terms. Instead, one may first calculate the Hartree and SOC contributions separately, in order to see whether these terms can lead to a momentum or energy-loss dependence of the scattering cross-section. Of particular interest is to see whether the spin asymmetry caused by SOC depends on these variables. In the following, we carefully analyze the Hartree and SOC matrix elements.

\subsection{Hartree matrix element}

The matrix element of the Hartree energy Eq.~(\ref{Eq:HartreeEnergy}) between the incoming and scattered wavefunctions given by Eq.~(\ref{Eq:ScatteringStates}) is (see Appendix~\ref{App:CoulombPotential})
\begin{multline}\label{Eq:HartreeMatrixElement}
	\langle n,s|\hat{V}_{\mathrm{H}}|m,i\rangle=\frac{eN_sN_i}{2\epsilon_0q}
	\left[\frac{\vec{s}^{\dagger}\!\cdot\vec{i}}{q+iq_z^-}
	+\frac{\vec{s}^{\dagger}\!\cdot(\mathcal{R}\vec{i})}{q+iq_z^+}\right.\\ \left.
	+\frac{(\mathcal{R}\vec{s})^{\dagger}\!\cdot\vec{i}}{q-iq_z^+}
	+\frac{(\mathcal{R}\vec{s})^{\dagger}\!\cdot(\mathcal{R}\vec{i})}{q-iq_z^-}\right]P_{nm}(\vec{q}),
\end{multline}
where $q_z^{\pm}=k_s^z\pm k_i^z$ and $q=|\vec{q}|$ is the norm of the two-dimensional vector $\vec{q}=\vec{k}_s-\vec{k}_i$. The quantity $P_{nm}(\vec{q})$ is given by $P_{nm}(\vec{q})=\int_{-\infty}^0 dz\,\langle n|\hat{\rho}(\vec{q},z)|m\rangle e^{-q|z|}$, where $\hat{\rho}(\vec{q},z)$ is the two-dimensional Fourier transform of $\hat{\rho}(\vec{R})\equiv\hat{\rho}(\vec{r},z)$.
The scattering cross-section including only the Hartree term can now be calculated by inserting Eq.~(\ref{Eq:HartreeMatrixElement}) into Eq.~(\ref{Eq:crossgeneral}). We will discuss the consequences of the reflection matrix on the Hartree cross-section in Sec.~\ref{Sec:Discussions}.

\subsection{Spin--orbit matrix element}

In a similar way one can calculate the matrix element of the SOC term. We first express it for an unspecified electric field and, in a second step, we specialize to the Coulomb field $-\vec{\nabla}\hat{V}_{\mathrm{H}}(\vec{R})/e$, which provides the relativistic correction to the Hartree scattering. For convenience, Eq.~(\ref{Eq:SOCEnergy}) is written in the from
\begin{equation}
	\hat{V}_{\mathrm{SOC}}=e\alpha\sum_{\nu\in\{x,y,z\}}\sigma_{\nu}\hat{O}_{\nu},
\end{equation}
where $\alpha=\hbar^2/(4m_e^2c^2)$ and the operators $\hat{O}_{\nu}$ only act on the orbital part of the electron wavefunction, for instance $\hat{O}_x=-i(\hat{\vec{E}}\times\vec{\nabla})_x=-i\hat{E}_y\partial/\partial z+i\hat{E}_z\partial/\partial y$ and similarly for $\hat{O}_y$ and $\hat{O}_z$. For the electron wavefunction Eq.~(\ref{Eq:ScatteringStates}), one sees that the derivatives $\partial/\partial x$ and $\partial/\partial y$ are equivalent to multiplication by $ik_i^x$ and $ik_i^y$, respectively. Furthermore, when calculating the matrix element $\langle s|\hat{V}_{\mathrm{SOC}}|i\rangle$, the in-plane integral on $\vec{r}=(x,y)$ simply yields the Fourier components of the electric field at the wavevector $\vec{q}$. One can, therefore, define the operators $\hat{O}_{\nu}(z)$ for the remaining integral on $z$:
\begin{subequations}\begin{align}
	\hat{O}_x(z)&=-i\hat{E}_y(\vec{q},z)\frac{\partial}{\partial z}-k_i^y\hat{E}_z(\vec{q},z)\\
	\hat{O}_y(z)&=k_i^x\hat{E}_z(\vec{q},z)+i\hat{E}_x(\vec{q},z)\frac{\partial}{\partial z}\\
	\hat{O}_z(z)&=k_i^y\hat{E}_x(\vec{q},z)-k_i^x\hat{E}_y(\vec{q},z).
\end{align}\end{subequations}
It follows that the SOC matrix element is given by 
\begin{widetext}
\begin{multline}\label{Eq:VSO0}
	\langle s|\hat{V}_{\mathrm{SOC}}|i\rangle=e\alpha N_sN_i\sum_{\nu}\left[
	\vec{s}^{\dagger}\!\cdot\sigma_{\nu}\vec{i}
	\int_0^{\infty}dz\,e^{-ik_s^zz}\hat{O}_{\nu}(z)e^{ik_i^zz}\Theta(z)
	+\vec{s}^{\dagger}\!\cdot\sigma_{\nu}(\mathcal{R}\vec{i})
	\int_0^{\infty}dz\,e^{-ik_s^zz}\hat{O}_{\nu}(z)e^{-ik_i^zz}\Theta(z)\right.\\
	\left.+(\mathcal{R}\vec{s})^{\dagger}\!\cdot\sigma_{\nu}\vec{i}
	\int_0^{\infty}dz\,e^{ik_s^zz}\hat{O}_{\nu}(z)e^{ik_i^zz}\Theta(z)
	+(\mathcal{R}\vec{s})^{\dagger}\!\cdot\sigma_{\nu}(\mathcal{R}\vec{i})
	\int_0^{\infty}dz\,e^{ik_s^zz}\hat{O}_{\nu}(z)e^{-ik_i^zz}\Theta(z)\right].
\end{multline}
In the above equation the summation is on $\nu=x, y, z$. Among them, the case of $\hat{O}_z$ is the simplest as it does not involve $\partial/\partial z$. The integrals may be expressed using the partial Fourier transform of the electric field along $z$:
\begin{equation}\label{Eq:VacuumElectricField}
	\hat{\vec{E}}(\vec{q},q_z)\equiv\int_0^{\infty}dz\,\hat{\vec{E}}(\vec{q},z)e^{-iq_zz}.
\end{equation}	
The first of the four terms involving $\hat{O}_z(z)$ in Eq.~(\ref{Eq:VSO0}) can thus be evaluated as
\begin{subequations}\begin{equation}
	\int_0^{\infty}dz\,e^{-ik_s^zz}\hat{O}_z(z)e^{ik_i^zz}\Theta(z)
	=k_i^y\hat{E}_x(\vec{q},q_z^-)-k_i^x\hat{E}_y(\vec{q},q_z^-).
\end{equation}
The three other terms yield the same result with $q_z^-$ replaced by $q_z^+$, $-q_z^+$, and $-q_z^-$, respectively. We continue with $\hat{O}_{x,y}(z)$, which yield additional contributions due to $\partial/\partial z$ acting on $\Theta(z)$:
\begin{align}
	\int_0^{\infty}dz\,e^{-ik_s^zz}\hat{O}_x(z)e^{ik_i^zz}\Theta(z)&=
	k_i^z\hat{E}_y(\vec{q},q_z^-)-\frac{i}{2}\hat{E}_y(\vec{q},z=0)-k_i^y\hat{E}_z(\vec{q},q_z^-)\\
	\int_0^{\infty}dz\,e^{-ik_s^zz}\hat{O}_y(z)e^{ik_i^zz}\Theta(z)&=
	k_i^x\hat{E}_z(\vec{q},q_z^-)-k_i^z\hat{E}_x(\vec{q},q_z^-)+\frac{i}{2}\hat{E}_x(\vec{q},z=0).
\end{align}\end{subequations}
The three subsequent terms have $(q_z^-,k_i^z)$ replaced by $(q_z^+,-k_i^z)$, $(-q_z^+,k_i^z)$, and $(-q_z^-,-k_i^z)$, respectively. Collecting everything, we arrive at the following expression for the SOC scattering matrix element:
\begin{subequations}\label{Eq:SOCMatrixElement}\begin{align}
	\langle s|\hat{V}_{\mathrm{SOC}}|i\rangle
	&=\langle s|\hat{V}^{(\mathrm{v})}_{\mathrm{SOC}}|i\rangle
	+\langle s|\hat{V}^{(\mathrm{s})}_{\mathrm{SOC}}|i\rangle\\
	\label{Eq:SOCMatrixElementb}
	\langle s|\hat{V}^{(\mathrm{v})}_{\mathrm{SOC}}|i\rangle
	&=e\alpha N_sN_i\left(\hat{\mathcal{E}}^+_{xyz}-\hat{\mathcal{E}}^+_{xzy}
	+\hat{\mathcal{E}}^+_{yzx}-\hat{\mathcal{E}}^+_{yxz}
	+\hat{\mathcal{E}}^-_{zxy}-\hat{\mathcal{E}}^-_{zyx}\right)\\
	\langle s|\hat{V}^{(\mathrm{s})}_{\mathrm{SOC}}|i\rangle
	&=-ie\alpha N_sN_i\frac{1}{2}\left(\vec{s}+\mathcal{R}\vec{s}\right)^{\dagger}\cdot
	\left[\sigma_x\hat{E}_y(\vec{q},z=0)-\sigma_y\hat{E}_x(\vec{q},z=0)\right]
	\left(\vec{i}+\mathcal{R}\vec{i}\right).
\end{align}
We have separated the ``vacuum'' terms proportional to $\hat{\vec{E}}(\vec{q},q_z)$, denoted by the superscript $(\mathrm{v})$, from the ``surface'' terms proportional to $\hat{\vec{E}}(\vec{q},z=0)$, denoted by the superscript $(\mathrm{s})$. In the ideal geometry with translation invariance in the $x$--$y$ plane, the electric field is parallel to $z$ and the latter terms drop. In Eq.~(\ref{Eq:SOCMatrixElementb}), the quantity $\hat{\mathcal{E}}$ is defined as
\begin{equation}
	\hat{\mathcal{E}}^{\pm}_{\gamma\nu\mu}=k_i^{\gamma}\left[
	\vec{s}^{\dagger}\!\cdot\sigma_{\nu}\vec{i}\,\hat{E}_{\mu}(\vec{q},q_z^-)
	\pm\vec{s}^{\dagger}\!\cdot\sigma_{\nu}(\mathcal{R}\vec{i})\hat{E}_{\mu}(\vec{q},q_z^+)
	+(\mathcal{R}\vec{s})^{\dagger}\!\cdot\sigma_{\nu}\vec{i}\,\hat{E}_{\mu}(\vec{q},-q_z^+)
	\pm(\mathcal{R}\vec{s})^{\dagger}\!\cdot\sigma_{\nu}(\mathcal{R}\vec{i})\hat{E}_{\mu}(\vec{q},-q_z^-)
	\right].
\end{equation}\end{subequations}
If the electric field is the one due to the Hartree potential, we have (see Appendix~\ref{App:ElectricField})
\begin{subequations}\label{Eq:ElectricField}\begin{align}
	\label{Eq:ElectricFielda}
	\langle n|\hat{\vec{E}}(\vec{q},q_z)|m\rangle
	&=\frac{\left(-iq_x,-iq_y,q\right)}{q+iq_z}\frac{P_{nm}(\vec{q})}{2\epsilon_0q}\\
	\label{Eq:ElectricFieldb}
	\langle n|\hat{E}_{x,y}(\vec{q},z=0)|m\rangle
	&=-iq_{x,y}\frac{P_{nm}(\vec{q})}{2\epsilon_0q}.
\end{align}\end{subequations}
We deduce the SOC matrix elements in this case:
\begin{subequations}\label{Eq:VSOC}\begin{align}
	\langle n,s|\hat{V}^{(\mathrm{v})}_{\mathrm{SOC}}|m,i\rangle&=\frac{e\alpha N_sN_i}{2\epsilon_0q}
	\left[q\left(\mathcal{F}^+_{xy}-\mathcal{F}^+_{yx}\right)
	-iq_x\left(\mathcal{F}^+_{yz}-\mathcal{F}^-_{zy}\right)
	+iq_y\left(\mathcal{F}^+_{xz}-\mathcal{F}^-_{zx}\right)\right]P_{nm}(\vec{q})\\
	\langle n,s|\hat{V}^{(\mathrm{s})}_{\mathrm{SOC}}|m,i\rangle&=\frac{e\alpha N_sN_i}{4\epsilon_0q}
	\left(\vec{s}+\mathcal{R}\vec{s}\right)^{\dagger}\cdot\left(q_x\sigma_y-q_y\sigma_x\right)
	\left(\vec{i}+\mathcal{R}\vec{i}\right)P_{nm}(\vec{q})\\
	\label{eq:F}
	\mathcal{F}^{\pm}_{\gamma\nu}&=k_i^{\gamma}\left[
	\frac{\vec{s}^{\dagger}\!\cdot\sigma_{\nu}\vec{i}}{q+iq_z^-}
	\pm\frac{\vec{s}^{\dagger}\!\cdot\sigma_{\nu}(\mathcal{R}\vec{i})}{q+iq_z^+}
	+\frac{(\mathcal{R}\vec{s})^{\dagger}\!\cdot\sigma_{\nu}\vec{i}}{q-iq_z^+}
	\pm\frac{(\mathcal{R}\vec{s})^{\dagger}\!\cdot\sigma_{\nu}(\mathcal{R}\vec{i})}{q-iq_z^-}\right].
\end{align}\end{subequations}
\end{widetext}
In Sec.~\ref{Sec:Discussions}, we will discuss the importance of the SOC matrix elements introduced in Eq.~(\ref{Eq:VSOC}).

\section{Discussion}\label{Sec:Discussions}

The spin-resolved Hartree cross-section can be calculated by inserting Eq.~(\ref{Eq:HartreeMatrixElement}) into Eq.~(\ref{Eq:crossgeneral}). It reads as
\begin{multline}\label{Eq:sigmaH}
	\frac{d^2S_{\mathrm{H}}}{d\Omega d\hbar\omega}=\left(\frac{2\pi}{\hbar}\right)^4m_e^2\frac{k_s}{k_i}
	\left(\frac{e}{2\epsilon_0q}\right)^2(N_sN_i)^2\\
	\times\left|\frac{\vec{s}^{\dagger}\!\cdot\vec{i}}{q+iq_z^-}
	+\frac{\vec{s}^{\dagger}\!\cdot(\mathcal{R}\vec{i})}{q+iq_z^+}
	+\frac{(\mathcal{R}\vec{s})^{\dagger}\!\cdot\vec{i}}{q-iq_z^+}
	+\frac{(\mathcal{R}\vec{s})^{\dagger}\!\cdot(\mathcal{R}\vec{i})}{q-iq_z^-}\right|^2\\
	\times\int_{-\infty}^0dzdz'\,\mathcal{S}(\vec{q},z,z',\omega)e^{-q|z+z'|},
\end{multline}
where $\mathcal{S}(\vec{q},z,z',\omega)$ is the spectral function
\begin{multline}\label{Eq:Spectralfunc}
	\mathcal{S}(\vec{q},z,z',\omega)=\frac{1}{Z}\sum_{mn}e^{-E_m/k_{\mathrm{B}}T}\\
	\times\langle m|\hat{\rho}(-\vec{q},z)|n\rangle\langle n|\hat{\rho}(\vec{q},z')|m\rangle
	\delta(\hbar\omega+E_m-E_n).
\end{multline}
The cross-section (scattering intensity) provides direct information on the frequency and momentum dependent dynamic (charge) response of the sample. It is important to notice that this quantity is probed over a depth which scales with $1/q$. Given the rather high momentum resolution of the experiments (on the order of 0.03~\AA$^{-1}$) this probing depth can be rather large. This means that the information on the dynamic response is not only restricted to the surface region but also several tens of nanometers below the surface. Owing to the long-range nature of the Coulomb interaction, this is not surprising. The penetration depth and the mean free path of the involved electrons are not relevant in this context.

If the reflection coefficient does not depend on the spin of the electrons and is real, the reflection matrix has the form $\mathcal{R}=R\sigma_0$ with $R$ a real constant. In this case, the spin-resolved Hartree cross-section becomes
\begin{multline}
	\frac{d^2S_{\mathrm{H}}}{d\Omega d\hbar\omega}\approx\left(\frac{2\pi}{\hbar}\right)^4\frac{k_s}{k_i}
	\left(\frac{m_ee}{\epsilon_0}\right)^2(N_sN_i)^2
	\frac{R^2|\vec{s}^{\dagger}\!\cdot\vec{i}|^2}{[q^2+(q_z^+)^2]^2}\\
	\times\int_{-\infty}^0dzdz'\,\mathcal{S}(\vec{q},z,z',\omega)e^{-q|z+z'|}.
\end{multline}
Note that while expressing $|\cdots|^2$ in Eq.~(\ref{Eq:sigmaH}) we kept only the middle terms. This is justified by the fact that for nearly specular reflection, we have $q\approx 0$ and $k_s^z\approx -k_i^z$, such that $|q_z^+|\ll|q_z^-|$. Consequently, the terms proportional to $(q\pm iq_z^+)^{-1}$ are dominant relative to those proportional to $(q\pm iq_z^-)^{-1}$. The same simplification is used in Refs.~\cite{Evans1972, Mills1975, Vig2017}. The factor $|\vec{s}^{\dagger}\!\cdot\vec{i}|^2$ expresses the conservation of spin. It can be replaced by unity for a non-spin-resolved experiment. The quantity $q_z^+$ can be expressed as $q_z^+=(v_{\parallel}q-\omega)/v_{\perp}$ (here $v_{\parallel}$ and $v_{\perp}$ denote the components of the incident electron's velocity parallel and perpendicular to the surface, respectively). Hence the results are exactly the same as those of Refs.~\cite{Evans1972, Mills1975, Vig2017}. Obviously, no spin asymmetry is expected in this case.

Likewise, the SOC cross-section can be expressed by inserting the matrix elements given by Eq.~(\ref{Eq:VSOC}). 
On a crystalline surface, the electric field has the periodicity of the lattice in the $x$--$y$ plane with Fourier amplitudes only at the reciprocal-lattice vectors. The $x$ and $y$ components of the electric field have vanishing amplitude at the center of the two-dimensional Brillouin zone, unlike the $z$ component. Hence at low $q$ only the $z$ component of the electric field contributes significantly to the scattering. The surface terms in Eq.~(\ref{Eq:VSOC}), that only stem from the $x$ and $y$ components of the field, can therefore be neglected. The same holds for disordered surfaces, as the low-$q$ scattering probes the spatial average of the electric field, which by symmetry must be oriented along $z$.

Ignoring the surface terms for further evaluation of the SOC cross-section leads to
\begin{subequations}\label{Eq:sigmaSOC}\begin{multline}
	\frac{d^2S_{\mathrm{SOC}}}{d\Omega d\hbar\omega}=\left(\frac{2\pi}{\hbar}\right)^4m_e^2\frac{k_s}{k_i}
	\left(\frac{e}{2\epsilon_0q}\right)^2(N_sN_i)^2\alpha^2\\
	\times\left|q\left(\mathcal{F}^+_{xy}-\mathcal{F}^+_{yx}\right)
	-iq_x\left(\mathcal{F}^+_{yz}-\mathcal{F}^-_{zy}\right)
	+iq_y\left(\mathcal{F}^+_{xz}-\mathcal{F}^-_{zx}\right)\right|^2\\
	\times\int_{-\infty}^0dzdz'\,\mathcal{S}(\vec{q},z,z',\omega)e^{-q|z+z'|}.
\end{multline}
If one only keeps the terms proportional to $\mathcal{R}$ in $|\cdots|^2$, similar to the case of the Hartree cross-section, then
\begin{equation}
	\mathcal{F}^{\pm}_{\gamma\nu}\approx k_i^{\gamma}\left[
	\pm\frac{\vec{s}^{\dagger}\!\cdot\sigma_{\nu}(\mathcal{R}\vec{i})}{q+iq_z^+}
	+\frac{(\mathcal{R}\vec{s})^{\dagger}\!\cdot\sigma_{\nu}\vec{i}}{q-iq_z^+}\right].
\end{equation}\end{subequations}

Now one can calculate the spin asymmetry caused by the SOC term. Since in the usual SPHREELS experiments only a spin-polarized incident beam is used, the scattering intensity is a sum over the partial intensities of both spin states after the scattering. We define the intensity $I_{|+\rangle}$ ($I_{|-\rangle}$) as the sum of the partial intensities when an electron of $|+\rangle$ ($|-\rangle$) initial state is scattered to either a $|+\rangle$ or a $|-\rangle$ final state. If the reflection matrix is independent of the spin, $\mathcal{R}=R\sigma_0$, we find that the asymmetry vanishes \footnote{A small asymmetry of order $(q_z^+/q_z^-)(R-1/R)(k_i^xq_y-k_i^yq_x/(k_i^zq)$ subsists if all terms in Eq.~(\ref{eq:F}) are retained in the calculation.}. This is an important result and indicates that the asymmetry is almost entirely defined by the reflection matrix. This means that the electrons inelastically scattered from a surface with a large SOC follow the same spin asymmetry as that of the elastically scattered ones. We emphasize that this is true only for small energy losses ($\hbar\omega \ll E_i$).

The asymmetry of elastically scattered electrons from surfaces with a large SOC has been investigated in great details \cite{Kirschner1985, Feder1986, Wang1979, McRae1979, McRae1981, Feder1986, Yu2007, Burgbacher2013, Vasilyev2015}. It has been realized that the presence of the surface barrier at the surfaces cannot create a noticeable spin asymmetry \cite{Kirschner1985, Feder1986}. This is due to the fact that the potential gradients caused by such potential profiles and the associated electric fields are usually too small, compared to the potential gradients in the vicinity of the nuclei. The presence of the SOC and the surface barrier has indirect consequences on the electron transmissivity and reflectivity. One, therefore, may assume that the electron reflectivity is spin dependent \cite{McRae1981, Kirschner1985, Feder1986}. This consideration can drastically simplify the situation. Based on Eq.~(\ref{Eq:sigmaSOC}) the scattering cross-section by the SOC term is scaled by a factor $\sim(\alpha q k_i)^2$ and, therefore, might be neglected in the total intensity for slow electrons. In this case, the Hartree cross-section considering spin-dependent reflection coefficient may be used to describe the results of SPHREELS experiments. In order to show that such an assumption is valid, we calculate the spin asymmetry caused by this term. Starting from Eq.~(\ref{Eq:sigmaH}) and assuming that $\mathcal{R}$ is a Hermitian matrix, the spin asymmetry is then given by
\begin{equation}\label{Eq:asymmetry1}
	\frac{I_{|+\rangle}-I_{|-\rangle}}{I_{|+\rangle}+I_{|-\rangle}}
	=\frac{|\mathcal{R}_{++}|^2-|\mathcal{R}_{--}|^2}
	{|\mathcal{R}_{++}|^2+|\mathcal{R}_{+-}|^2+|\mathcal{R}_{--}|^2+|\mathcal{R}_{-+}|^2}.
\end{equation}
The real values $|\mathcal{R}_{\sigma\sigma'}|^2$  can be measured by elastic reflectivity measurements. They represent the partial intensities of the scattered electrons when the spin of the incoming and outgoing beam is of $\sigma$ and $\sigma'$ character, respectively. In the so-called \emph{complete} experiment, where a spin-polarized beam is used as the source and the detection is also spin resolved, all the four possible partial intensities are known \cite{Kirschner1985a}. Such experiments can be designed with a high momentum resolution but usually suffer from a poor energy resolution, due to the inefficiency of the spin detectors for the inelastic part of the scattering \cite{Vasilyev2016}. Hence, in the usual SPHREELS experiments only the incoming beam is spin polarized and the detection is not spin-resolved \cite{Kirschner1984, Zakeri2013, Zakeri2014, Zakeri2021}. Therefore, the partial intensities of both spin characters in the final state are added up. This means that in the experiment the values of $|\mathcal{R}_{|+\rangle}|^2=|\mathcal{R}_{++}|^2+|\mathcal{R}_{+-}|^2$ and $|\mathcal{R}_{|-\rangle}|^2=|\mathcal{R}_{--}|^2+|\mathcal{R}_{-+}|^2$ are measured and Eq.~(\ref{Eq:asymmetry1}) can be simplified to
\begin{equation}\label{Eq:asymmetry2}
	\frac{I_{|+\rangle}-I_{|-\rangle}}{I_{|+\rangle}+I_{|-\rangle}}
	=\frac{|\mathcal{R}_{|+\rangle}|^2-|\mathcal{R}_{|-\rangle}|^2}
	{|\mathcal{R}_{|+\rangle}|^2+|\mathcal{R}_{|-\rangle}|^2}.
\end{equation}
Equation~(\ref{Eq:asymmetry2}) has an important implication. In the limit of $\hbar\omega \ll E_i$ the asymmetry of the inelastic scattering is independent of $\hbar\omega$ and $q$ and follows that of the elastic scattering. Hence, one can use the same numerical scheme originally developed for HREELS experiments \cite{Sunjic1971, Lucas1972, Lambin1990} to simulate the SPHREELS spectra.  The important observation is that one needs to consider the spin-dependent reflection coefficients. Recall that the reflection matrix is a $2 \times 2$ matrix with the reflection coefficient being its elements. In a general form these coefficients are complex entities and may be calculated using state of the art first-principles calculations. Such calculations require primarily the details of the geometrical and electronic structures of the sample. However, the square of the reflection coefficients are real values and can be measured by elastic reflectivity experiments. In this case one needs to measure the partial intensities of the elastically reflected beam for the incoming spin states $|+\rangle$  and $|-\rangle$ as a function of the incident electron beam by placing the detector in the specular geometry.

One more important point is that all the elements of the reflection matrix depend on the energy and angle of the incident beam \cite{Kirschner1979, Samarin2007}. Obviously, the asymmetry depends on these variables. For large values of energy losses the situation can become rather complicated. In such cases one may observe a strong energy-loss dependence of the spin asymmetry due to the fact that the reflection coefficients depend on the electron's energy just before scattering. If this energy is much different than the incident energy, the reflection coefficient is also different.

Very recently, high-resolution experiments have been reported on quantum materials with a large SOC and  interesting spin-dependent effects associated with SOC have been observed \cite{Zakeri2021,Zakeri2022a}. Our results shall provide the fundamental basis required to numerically calculate the spin-polarized spectra and understand the experimental results in details (an example may be found in Ref. \cite{Zakeri2022a}).

\section{Conclusion}

The successful theory by Evans and Mills describes the scattering cross-section of low-energy electrons from a nonmagnetic surface. The scattering process is based on the electrostatic Coulomb interaction of the electrons with the charge-density distribution of the sample. The theory describes most of the important features observed in the HREELS experiments, namely phonons, plasmons and any other collective excitation associated with the charge-density fluctuations. Aiming at a detailed understanding of the impact of SOC on the scattering cross-section of low-energy electrons scattered from a nonmagnetic surface, we extended the theory to the cases in which SOC is included in the formalism. In the presence of this term the scattering cross-section becomes spin-dependent. For a more appropriate description one has to also take into account that the elements of the reflection matrix are spin-dependent. We show that, if one assumes that the scattering cross-section caused by the SOC term is small, and the effect of SOC is to cause spin-dependent reflection coefficients, the spin asymmetry will be independent of the electron energy loss and the momentum transfer. Assuming a Hermitian reflection matrix, one can derive a simplified expression for the scattering cross-section, which can be used for numerical calculation of the spin-polarized spectra recorded in SPHREELS experiments.   In principle, one can use the same machinery developed and implemented by Lucas and \v{S}unji\'{c} \cite{Sunjic1971, Lucas1972, Lambin1990} and simulate the SPHREELS spectra \cite{Zakeri2022, Zakeri2022a}. The only difference in this case will be to use spin-dependent reflection coefficients. These quantities can be obtained by the elastic reflectivity measurements.

\section*{Acknowledgements}

Kh.Z.\ acknowledges funding from the Deutsche Forschungsgemeinschaft (DFG) through the Heisenberg Programme ZA 902/3-1 and ZA 902/6-1 and the DFG Grant Nos.\ ZA 902/5-1 and ZA 902/7-1. Kh.Z.\ thanks the Physikalisches Institut for hosting the group and providing the necessary infrastructure.

\bibliography{Refs}

\appendix

\section{Matrix element of the Hartree potential}\label{App:CoulombPotential}

Upon substitution of the two-dimensional Fourier representations of the charge-density operator, $\hat{\rho}(\vec{R})=(2\pi)^{-2}\int d^2k\,e^{i\vec{k}\cdot\vec{r}}\hat{\rho}(\vec{k},z)$, and of the Coulomb potential, $1/|\vec{R}|=(2\pi)^{-2}\int d^2k\,e^{i\vec{k}\cdot\vec{r}}(2\pi/k)e^{-k|z|}$, the matrix element of the Hartree operator, Eq.~(\ref{Eq:HartreeEnergy}), becomes
\begin{multline*}
	\langle s|\hat{V}_{\mathrm{H}}|i\rangle=\frac{e}{4\pi\epsilon_0}
	\int d\vec{R}\,\psi_s^*(\vec{R})\int d\vec{R}'
	\int\frac{d^2k}{(2\pi)^2}e^{i\vec{k}\cdot\vec{r}'}\\ \times\hat{\rho}(\vec{k},z')
	\int\frac{d^2k'}{(2\pi)^2}\,e^{i\vec{k}'\cdot(\vec{r}-\vec{r}')}
	\frac{2\pi}{k'}e^{-k'|z-z'|}\sigma_0\psi_i(\vec{R}).
\end{multline*}
For the wavefunctions given by Eq.~(\ref{Eq:ScatteringStates}), the in-plane integrals on $\vec{r}$ and $\vec{r}'$ yield $(2\pi)^4\delta(\vec{k}'-\vec{q})\delta(\vec{k}-\vec{k}')$, where the in-plane wavevector transfer is $\vec{q}=\vec{k}_s-\vec{k}_i$. Considering that the wavefunctions vanish if $z<0$ and that there are no charges for $z'>0$, the matrix element reduces to
\begin{multline*}
	\langle s|\hat{V}_{\mathrm{H}}|i\rangle=\frac{e}{2\epsilon_0q}
	\int_0^{\infty}dz\,\psi_s^*(\vec{0},z)\int_{-\infty}^0dz'\\
	\times\hat{\rho}(\vec{q},z')e^{-q|z-z'|}\sigma_0\psi_i(\vec{0},z).
\end{multline*}
The integral on $z$ can now be evaluated by means of the identity
\begin{equation}\label{Eq:partial}
	\int_0^{\infty} dz\,e^{-q|z-z'|}e^{iq_zz}=\frac{e^{-q|z'|}}{q-iq_z},
\end{equation}
which leads directly to Eq.~(\ref{Eq:HartreeMatrixElement}).

\section{Components of the electric field}\label{App:ElectricField}

The electric field due to the charge distribution $\hat{\rho}(\vec{R})$ is $\hat{\vec{E}}(\vec{R})=-\vec{\nabla}\hat{V}_{\mathrm{H}}(\vec{R})/e$, which after Fourier transform reads $\hat{E}_{x,y}(\vec{q},z)=-iq_{x,y}\hat{V}_{\mathrm{H}}(\vec{q},z)/e$ for the in-plane components and $\hat{E}_{z}(\vec{q},z)=-\partial/\partial z\hat{V}_{\mathrm{H}}(\vec{q},z)/e$ for the $z$ component. The $x$ and $y$ components are readily evaluated by introducing the Fourier representations of the charge density and Coulomb potential, like in Appendix~\ref{App:CoulombPotential}:
\begin{align*}
	\hat{E}_{x,y}(\vec{q},z)&=\frac{-iq_{x,y}}{4\pi\epsilon_0}\int d^2r\,e^{-i\vec{q}\cdot\vec{r}}
	\int d\vec{R}'\int\frac{d^2k}{(2\pi)^2}e^{i\vec{k}\cdot\vec{r}'}\\
	&\quad\times\hat{\rho}(\vec{k},z')\int\frac{d^2k'}{(2\pi)^2}\,e^{i\vec{k}'\cdot(\vec{r}-\vec{r}')}
	\frac{2\pi}{k'}e^{-k'|z-z'|}\\
	&=\frac{-iq_{x,y}}{2\epsilon_0q}\int_{-\infty}^0dz'\,\hat{\rho}(\vec{q},z')e^{-q|z-z'|}.
\end{align*}
This leads to Eq.~(\ref{Eq:ElectricFieldb}). We now perform the partial Fourier transform as in Eq.~(\ref{Eq:VacuumElectricField}), using Eq.~(\ref{Eq:partial}),
\begin{align*}
	\hat{E}_{x,y}(\vec{q},q_z)&=\int_0^{\infty}dz\,\hat{\vec{E}}(\vec{q},z)e^{-iq_zz}\\
	&=\frac{-iq_{x,y}}{2\epsilon_0q}\int_{-\infty}^0dz'\,\hat{\rho}(\vec{q},z')\frac{e^{-q|z'|}}{q+iq_z},
\end{align*}
which establishes Eq.~(\ref{Eq:ElectricFielda}) for the components $x$ and $y$. The expression of $\hat{E}_z(\vec{q},z)$ is the same as that of $\hat{E}_{x,y}(\vec{q},z)$ with $-iq_{x,y}$ replaced by $-\partial/\partial z$:
\begin{align*}
	\hat{E}_z(\vec{q},z)&=\frac{1}{2\epsilon_0q}\int_{-\infty}^0dz'\,\hat{\rho}(\vec{q},z')
	\left(-\frac{\partial}{\partial z}\right)e^{-q|z-z'|}\\
	&=\frac{1}{2\epsilon_0}\int_{-\infty}^0dz'\,\hat{\rho}(\vec{q},z')\mathrm{sign}(z-z')e^{-q|z-z'|}.
\end{align*}
The partial Fourier transform is performed using the identity
\begin{multline}
\int_0^{\infty}dz\,\mathrm{sign}(z-z')e^{-q|z-z'|}e^{-iq_zz}\\
	=\begin{cases}
	\displaystyle\frac{e^{-q|z'|}}{q+iq_z} & z'<0\\[1em]
	\displaystyle\frac{e^{-q|z'|}}{q-iq_z}-\frac{2iq_z}{q^2+q_z^2}e^{-iq_zz'} & z'>0,
	\end{cases}
\end{multline}
which proves Eq.~(\ref{Eq:ElectricFielda}) for the $z$ component.

\end{document}